\documentclass{aa} 

\usepackage{graphicx}
\usepackage{txfonts}
\usepackage{multicol}
\usepackage{multirow}
\usepackage{lscape}
\usepackage{adjustbox}
\usepackage{graphicx}
\usepackage{hyperref}   
\hypersetup{
    colorlinks=true,
    linkcolor=blue,
    filecolor=magenta,      
    urlcolor=cyan,
    pdftitle={Overleaf Example},
    pdfpagemode=FullScreen,
    }

\usepackage{amsmath,amssymb}
\usepackage{subfig}

\begin{document}

   \title{A failed wind candidate in NGC 3783 from the 2001 year campaign with Chandra/HETGS}


   \author{Chen Li
          \inst{1, 2},
          Jelle S. Kaastra\inst{1, 2},
          Liyi Gu\inst{2, 1},
          Daniele Rogantini\inst{3},
          Anna Jur\'{a}\v{n}ov\'{a}\inst{4}, \\
          Missagh Mehdipour\inst{5},
          Jelle de Plaa\inst{2}
          }

   \institute{Leiden Observatory, Leiden University, PO Box 9513, 2300 RA Leiden, the Netherlands\\
              \email{cli@strw.leidenuniv.nl}
         \and
             SRON Netherlands Institute for Space Research, Niels Bohrweg 4, 2333 CA Leiden, The Netherlands 
          \and
             Department of Astronomy and Astrophysics, The University of Chicago, Chicago, IL 60637
          \and
             MIT Kavli Institute for Astrophysics and Space Research, Massachusetts Institute of Technology, Cambridge, MA 02139, USA
          \and 
             Space Telescope Science Institute, 3700 San Martin Drive, Baltimore, MD 21218, USA
             }

   \date{}

 
  \abstract
  {
  We reanalyze the Chandra/HETGS observations of NGC 3783 from the campaign in the year 2001, identifying significant spectral variations in the Fe unresolved transition array (UTA) over timescales of weeks to months. 
  These changes correlate with a $1.4-2$ fold increase in the ionizing continuum and exceed $10 \, \sigma$ significance. 
  The variations primarily originate from a low-ionization state ($\rm log \xi = 1.65$) component of the warm absorber. 
  Time-dependent photoionization modelling confirms the sensitivity of this low-ionization component to continuum variations within the Fe UTA band. 
  Local fitting indicates a lower density limit of $>10^{12.3} \, \rm m^{-3}$ at $3 \, \sigma$ statistical uncertainty, with the component located within $0.27 \, \rm pc$. 
  Our findings suggest that this low-ionization component is a potential failed wind candidate.
  }

  \titlerunning{a new identified failed wind in NGC 3783}
  \authorrunning{C. Li et al.}
  \maketitle
  
%

\section{Introduction}\label{introduction}

NGC 3783, a nearby Seyfert 1 galaxy at redshift $\rm z=0.009730$ \citep{Theureau1998A&AS}, hosts one of the most luminous local AGNs, with bolometric luminosity of log $L_{\rm AGN} \sim  44.5 \, \rm erg \, \rm s^{-1}$ at a distance of $38.5 \rm \, Mpc$ \citep{Davies2015ApJ}.
The AGN is powered by a supermassive black hole with $M_{ \rm BH} = 2.82^{+1.55}_{-0.63} \times 10^{7} \, \rm M_{\sun}$  \citep{Bentz2021ApJ}. 
Extensive studies have focused on its ionized outflow, particularly the X-ray warm absorbers near the nucleus.
While near-infrared interferometry has partially resolved the spatial structure of the broad-line region (BLR) within the central parsec \citep{GRAVITY_Collaboration_2021A&A}, the lack of direct imaging limits precise distance measurements of ionized plasma, including the warm absorber. 
This limits our ability to accurately determine the kinetic power and mass outflow rate of these absorbers.

The ionization parameter
\begin{equation}\label{eq1}
\xi = \frac{L_{\rm ion}}{n_{\rm H} \, r^{2}} \,  ,  
\end{equation}
serves as a proxy for the outflow distance from the ionizing source, where $L_{\rm ion}$ is the ionizing luminosity over the 1-1000 Ryd range, $n_{\rm H}$ is the hydrogen number density, and $r$ is the distance from the ionizing source 
(\citealp{Tarter1969ApJ}, \citealp{Krolik1981ApJ}). 
By measuring $\xi$, $L_{\rm ion}$, and $n_{\rm H}$, we can indirectly estimate the distance of the outflow.

Two main approaches can be used to derive the density for AGN outflows.
The first relies on density-sensitive metastable spectral lines, which require high-quality, high-resolution spectra (often from time-averaged observations).
For example, using the spectral energy distribution (SED) of NGC 5548,  \cite{Mao2017A&A} studied density diagnostics for AGN outflows through metastable absorption lines of Be-, B-, and C-like ions, showing that different ions within the same isoelectronic sequence can cover a broad range of ionization parameters and densities.
This technique has been used successfully to constrain the density of the lower ionized gas in NGC 3783 using UV lines (\citealp{Gabel2005ApJ}).

As a another approach, the spectral-timing method uses time-resolved spectra to analyze plasma responses to fluctuations in ionizing luminosity (\citealp{Kaastra2012A&A}, \citealp{Silva2016A&A}, \citealp{Juranova2022MNRAS}). 
Time-dependent photoionization modeling provides a comprehensive framework, simultaneously solving for ion concentration, heating, and cooling evolution in response to SED and AGN variability (\citealp{Rogantini2022ApJ}, \citealp{Sadaula2023ApJ}). 
The evolution of plasma state, indicated by average charge over time, is driven by the relationship between the variability timescale ($t_{\rm var}$) and the recombination timescale ($t_{\rm rec}$), where $t_{\rm rec}$ is inversely proportional to plasma density. 
When $t_{\rm var}$ and $t_{\rm rec}$ are comparable, measurable lag timescales emerge between ionizing luminosity changes and plasma state variations, given adequate sampling and photon counts in individual spectra  (\citealp{Li2023A&A}).

Variability in spectral lines correlated with flux changes has been observed in high signal-to-noise, high-resolution absorption spectroscopic studies (\citealp{Netzer2003ApJ}; \citealp{Krongold2005ApJ}). Analyzing the 900 ks Chandra/HETG spectrum of NGC 3783, \citealp{Kaspi2002ApJ} provided a comprehensive characterization of the absorption spectrum, which included several significant iron features, such as the L-shell lines from \ion{Fe}{XVII} to \ion{Fe}{XXIV}, as well as the Unresolved Transition Array (UTA) of M-shell lines. The distinct Fe UTA structure has also been detected in XMM-Newton RGS spectra (\citealp{Behar2001ApJ}; \citealp{Blustin2002A&A}).
Various photoionization models have been applied to analyze the X-ray absorption spectra of NGC 3783, employing different configurations: two ionized components (\citealp{Blustin2002A&A}; \citealp{Krongold2003ApJ}; \citealp{Krongold2005ApJ}), three components (\citealp{Netzer2003ApJ}), five components (\citealp{Ballhausen2023ApJ}), and nine components (\citealp{Mehdipour2017A&A}; \citealp{Kaastra2018A&A}; \citealp{Mao2019A&A}). These models typically assumed photoionization equilibrium. However, \cite{Gu2023A&A} introduced time-dependent effects within a nine-component photoionization model, constraining the warm absorber’s density to the range $10^{10} - 10^{13} \rm m^{-3}$.

Using the new time-dependent photoionization model, \texttt{tpho} (\citealp{Rogantini2022ApJ}), in the SPEX code (\citealp{Kaastra2024spex}), we perform a self-consistent calculation of the full time-dependent ionization state for all ionic species, generating synthetic transmission spectra based on the observed light curve.

The structure of the paper is as follows. 
Section \ref{Data_reduction} presents the data reduction process for the Chandra/HETGS observations, incorporating recent calibration updates and displaying the resulting ratio spectrum. 
We also use the long-term RXTE light curve to calibrate the six Chandra observations to RXTE’s flux level.
Section \ref{Method} describes our application of the \texttt{tpho} model to analyze the Chandra grating data of NGC 3783.
Section \ref{results} summarizes the findings from our analysis.
Finally, Section \ref{discussion} discusses the results and potential model dependencies.

\section{Data Reduction}\label{Data_reduction}

We used the $\rm CIAO \, v4.15$ software and calibration database $\rm (CALDB) \, v4.10$ to reprocess the Chandra/HETGS data.
The \texttt{chandra\textunderscore{\texttt{repro}}}  script was employed to screen the data and generate spectral files for each observation. We combined the +1st and -1st orders of the medium-energy grating (MEG) spectra using the CIAO tool \texttt{combine}\textunderscore{\texttt{grating}}\textunderscore{\texttt{spectra}}, along with the associated response files, to produce an averaged MEG spectrum for each observation, with Gehrel's errors applied.

Next, we used the HEASoft \texttt{mathpha} tool to convert Gehrel's errors to Poisson errors and the spextools \texttt{trafo} utility to transform the OGIP spectral format into the SPEX data input format.

\begin{figure*}
    \centering
    \includegraphics[width=1.0\linewidth]{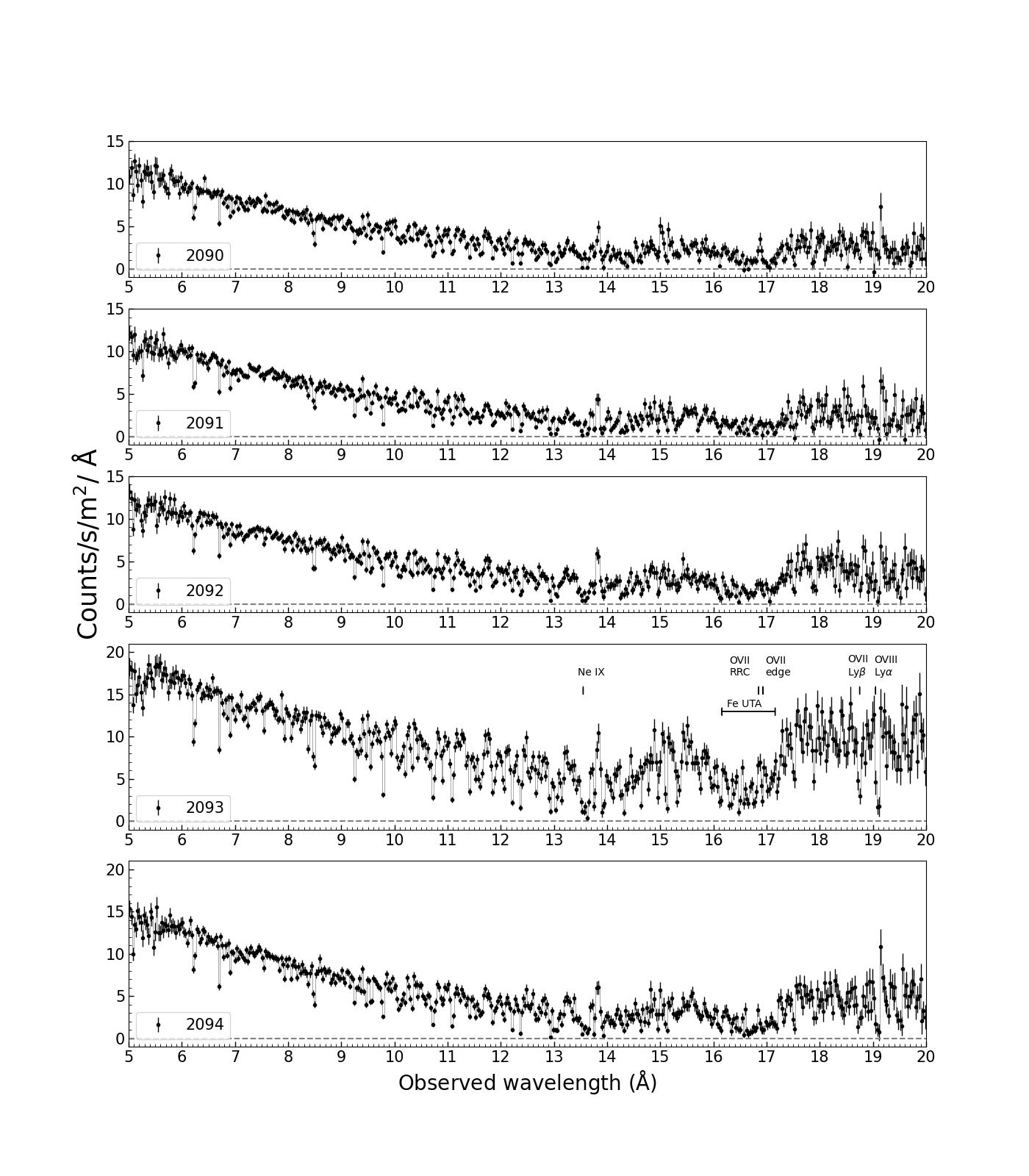}
    \caption{Chandra/HETGS observations in 2001 binned at $0.03 \, \rm \AA$.
    }
    \label{fig:chandra 6 obs}
\end{figure*}

To minimize systematic uncertainties, we utilize only the first-order Chandra/MEG spectrum, as the second and third orders are too faint to significantly contribute to the photon count. Fig. \ref{fig:chandra 6 obs} displays the resulting spectra for the five observations from the 2001 campaign, with the Fe UTA feature clearly visible in the high-flux observation (ObsID 2093).

To investigate potential line variability, we calculate ratio spectra by comparing the high-flux state (ObsID 2093) to four lower-flux states (ObsIDs 2090, 2091, 2092, and 2094),  which were observed within approximately one to three months of the high-flux state.

\begin{figure*}
    \centering
    \includegraphics[width=1.0\linewidth]{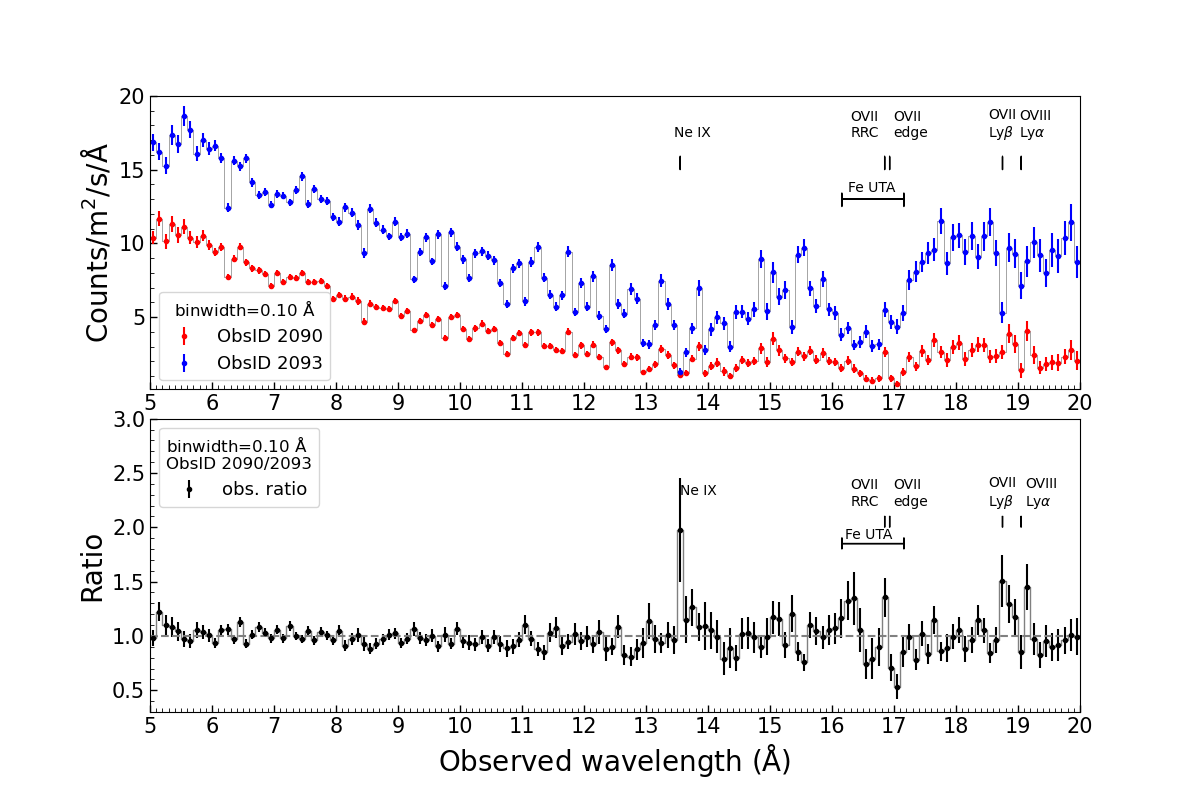}
    \caption{
    Top: low flux state ObsID 2090 and high flux state ObsID 2093 spectra binned to $0.1 \rm \, \AA$.
    Bottom: The low flux state ObsID 2090 is scaled up by a factor of 2.21, and the ratio spectrum is defined by the low flux state over the high flux state ObsID 2093.
    }
    \label{fig:2090-2093_ratio_0.1ang_5-20ang}
\end{figure*}

To account for the effect of continuum variation, we normalize the low-flux data to the high-flux state based on the average flux in the 9 – 14 $\AA$ band. 
The resulting ratio spectrum between ObsID 2090 and ObsID 2093 is presented in Fig. \ref{fig:2090-2093_ratio_0.1ang_5-20ang}.

\begin{table}
\caption{Observation log.
$t_{\rm exp}$: exposure time of Chandra observation.
$t_{\rm gap}$: the time gap between two adjacent Chandra observations.
The fifth column shows the averaged Chandra flux for each observation.
RXTE flux (last column) in the $2$ - $60$ $\rm ke V$ band.
}
\label{table: K05 and N03}
\begin{adjustbox}{max width=0.5\textwidth}
\begin{tabular}{l|l|l|l|l|l}
\hline
ID & start time& $t_{\rm exp}$&  $t_{\rm gap}$& Chandra ave. flux&  RXTE flux\\
&    (UTC)&    &    &    (5-20)$\rm \AA$&   (2-60 $\rm keV$)  \\
  &     yyyy-mm-dd&  (ks)&   (days)& $(\rm cts/s/m^{2}/ \AA$)&   (cts/s/PCU)    \\
\hline
373& 2000-01-20& 56& &  6.26&  9.507 \\
\hline
2090& 2001-02-24& 166&  369d& 4.10& 6.505 \\
\hline
2091& 2001-02-27& 169& 3d&  4.01&  6.405 \\
\hline
2092& 2001-03-10& 165&  11d&  4.81&  7.118 \\
\hline
2093& 2001-03-31& 166&  21d&  8.98&  11.58 \\
\hline
2094& 2001-06-26& 166&  108d&  5.59&  8.389 \\
\hline
\end{tabular}
\end{adjustbox}
\end{table}


\begin{figure*}
    \centering
    \includegraphics[width=1.0\linewidth]{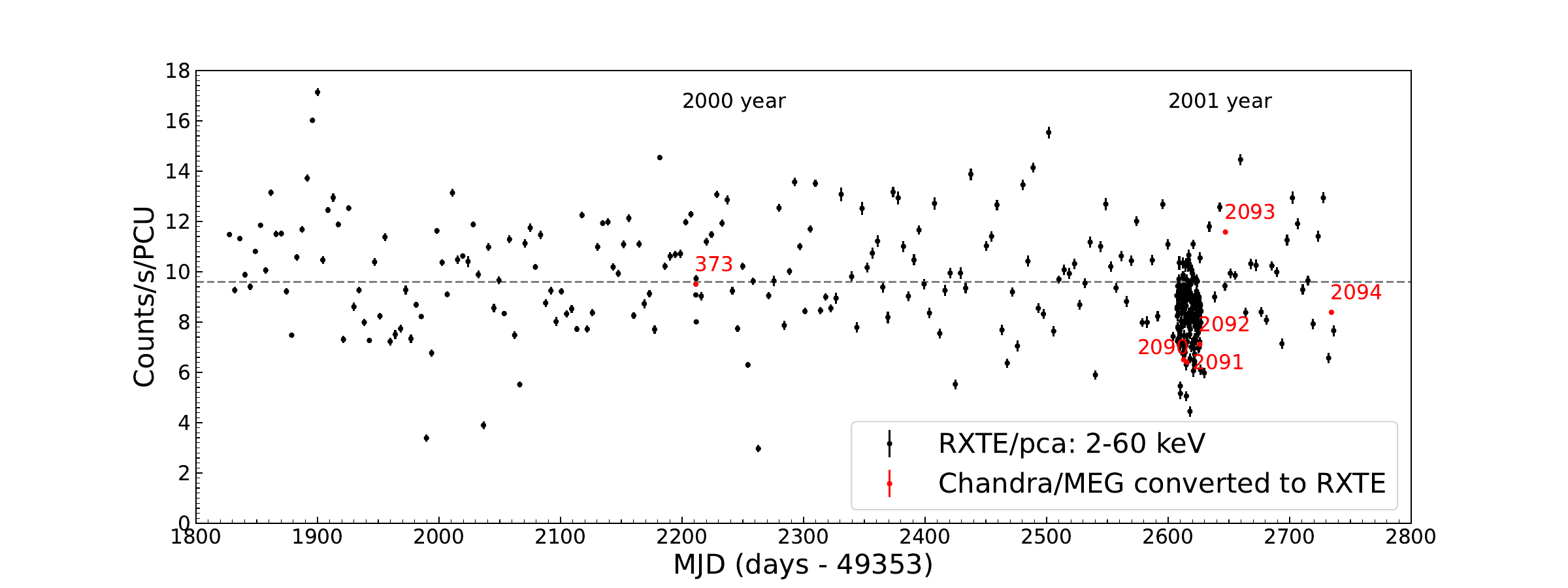}
    \includegraphics[width=1.0\linewidth]{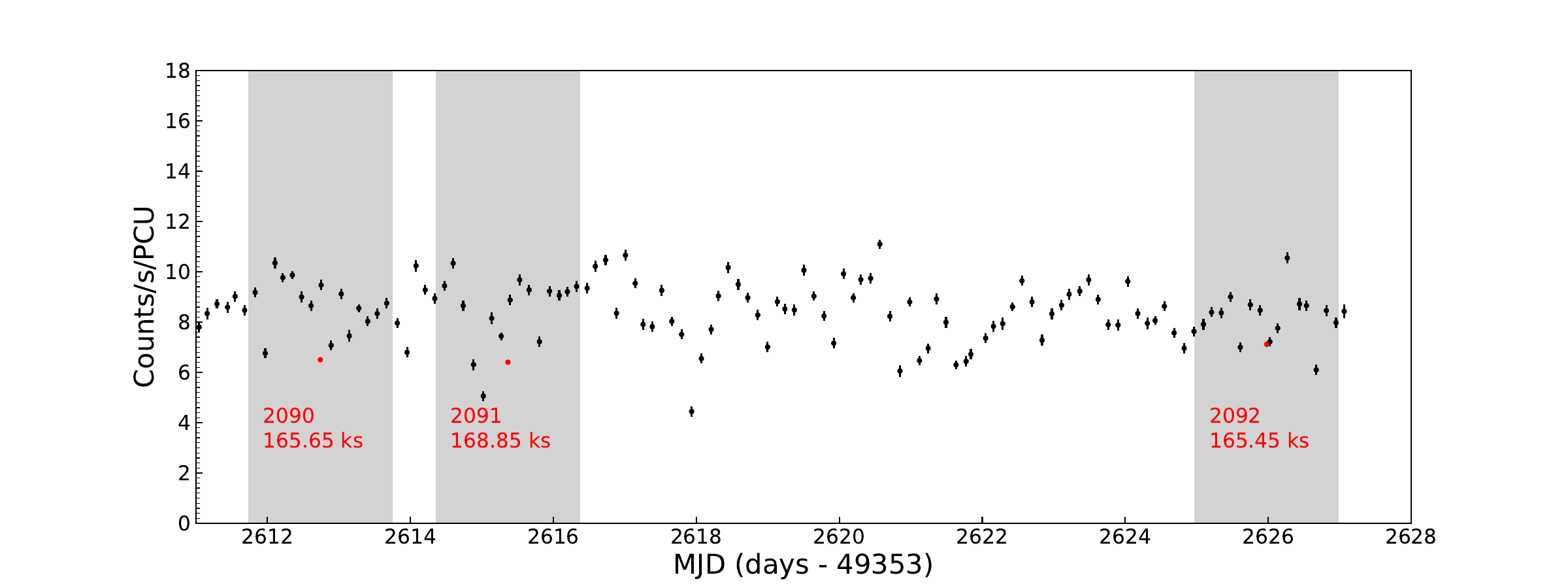}
    \includegraphics[width=1.0\linewidth]{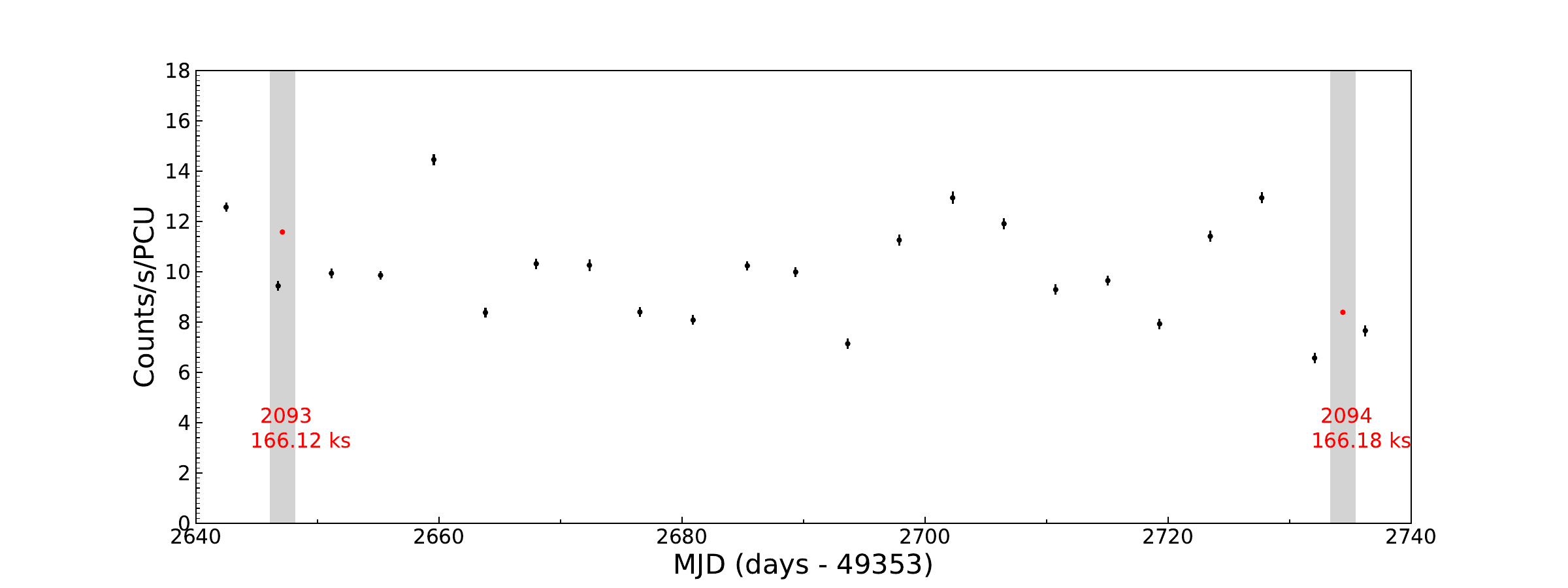}
    \caption{RXTE light curve of NGC 3783 together with the epochs of 6 Chandra observations. 
    Top panel: the horizontal line represents the average flux level over 1.5 years of RXTE observations.
    Light gray shade bars represent the epochs of the Chandra observations.
    }
    \label{fig:combination light curve for see}
\end{figure*}

To construct a continuous light curve, we combine RXTE/PCA and Chandra/HETGS data, as shown in Fig. \ref{fig:combination light curve for see}. 
Chandra/HETGS fluxes are converted to the equivalent RXTE 2-60 keV flux using WebPIMMS, assuming a power-law continuum and a Galactic absorption column density of $8.7 \times 10^{20} \, \rm cm^{-2}$ (\citealp{Alloin1995A&A}). 
The converted flux levels are provided in Table \ref{table: K05 and N03}. 
The Chandra observations analyzed in this study are marked in red in Fig. \ref{fig:combination light curve for see}, plotted at the midpoint of each exposure.
For reference to long-term variability, we also include the observation taken in 2000 (ObsID 373) to illustration the long-term timescale.

\section{Method}\label{Method}

\begin{figure}
    \centering
    \includegraphics[width=1.0\linewidth]{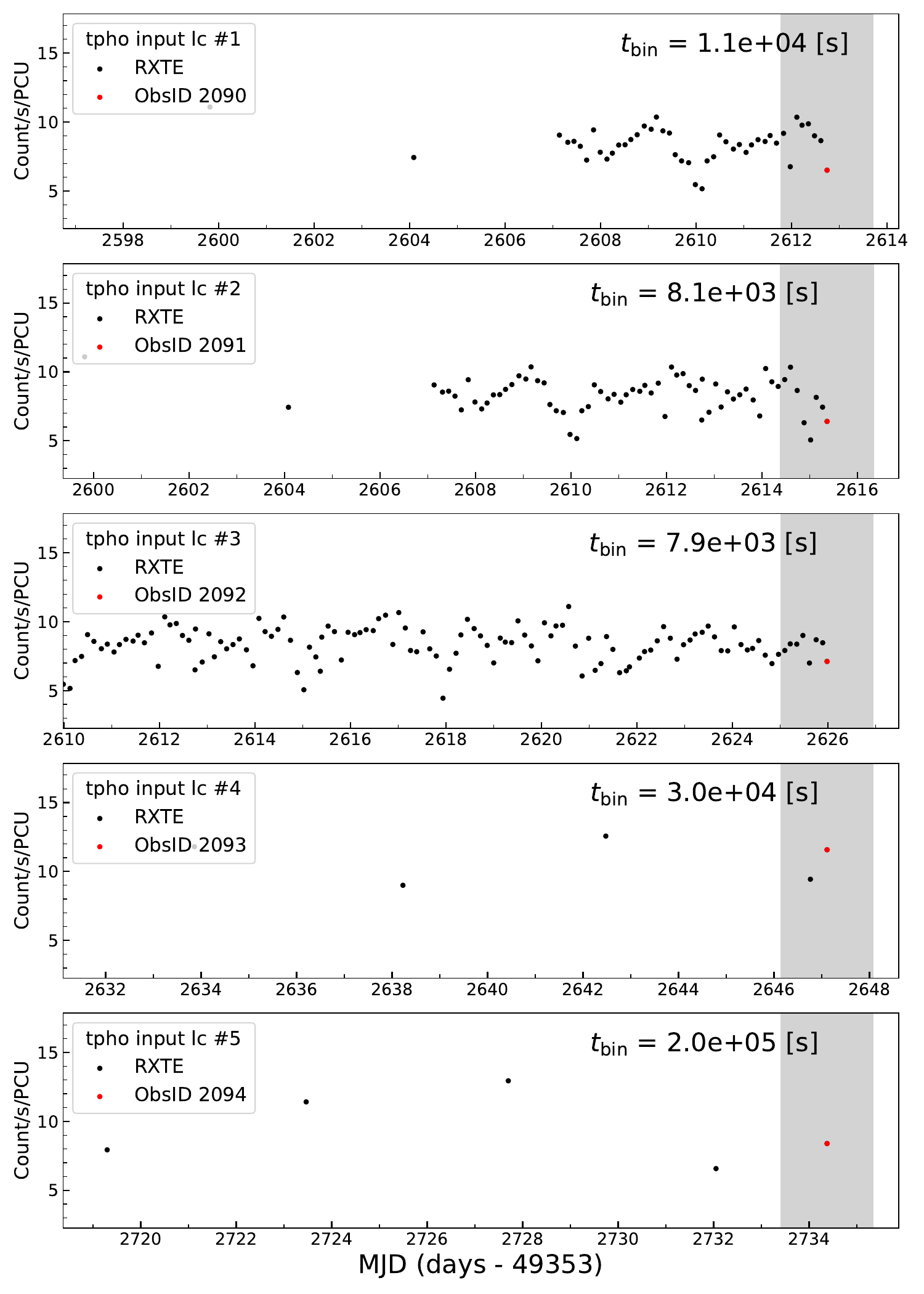}
    \caption{RXTE light curve used as input for each \texttt{tpho} component calculation with the same starting data point from 1800 MJD. 
    We show here the part of light curve within 14 days before the Chandra observation (red dot) corresponding to each component.
    The markers are similar to Fig. \ref{fig:combination light curve for see}.
    $t_{\rm bin}$ (time resolution) in unit of second was defined by the time difference between the Chandra observation and the last RXTE observation before it.}
    \label{fig:6 lc for tpho}
\end{figure}

To investigate the spectral variability, we employ the \texttt{tpho}  model (\citealp{Rogantini2022ApJ}) of SPEX (v3.08.01; \citealp{Kaastra1996}; \citealp{Kaastra2024spex}). 
Following \cite{Li2023A&A}, the initial unobscured SED is derived from the average broadband spectrum presented by \cite{Mehdipour2017A&A} with no assumed changes in SED shape over time.
The nine warm absorber components identified by \cite{Mao2019A&A} are implemented within \texttt{tpho}, with their equilibrium properties serving as initial conditions. 
To capture the effects of long-term variability, we trace the evolution of each warm absorber component, extending sufficiently far back in time, guided by the up to weeks of RXTE monitoring preceding the Chandra observations are shown in
 Fig. \ref{fig:6 lc for tpho}.

Each component is exposed to the same unobscured SED, with light curves varying according to the observed data. This approach enables us to assess how flux amplitude influences the resulting absorption spectra. The simulated transmissions are then cosmologically redshifted and convolved with the MEG response function. Finally, we derive the model ratio between low- and high-flux states from the simulated transmissions, with each state normalized separately to its respective continuum level.

%


\begin{figure}
    \centering
    \includegraphics[width=1.0\linewidth]{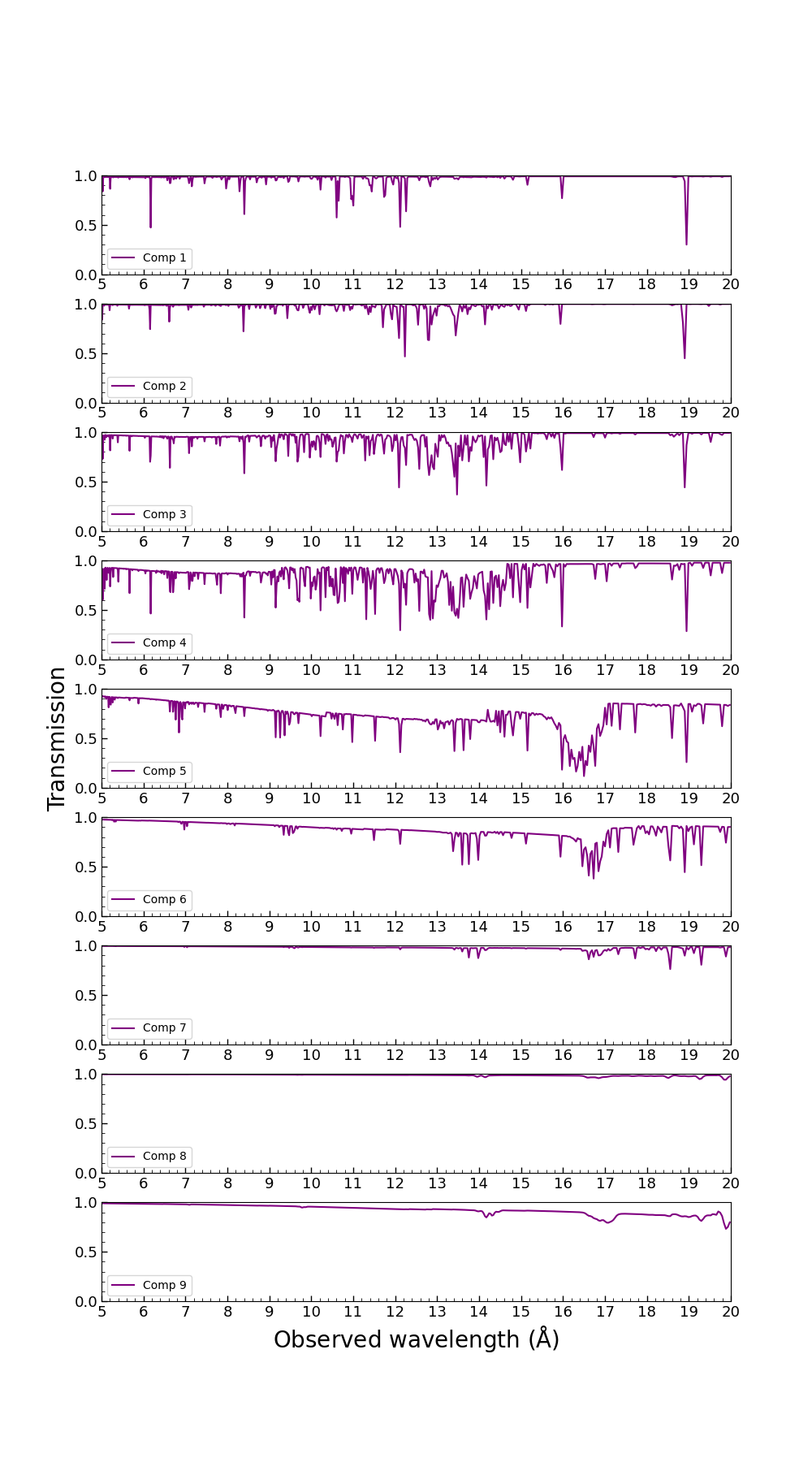}
    \caption{Transmissions of the nine warm absorber components of NGC 3783 computed with the \texttt{pion} model based on the parameters listed in \cite{Mao2019A&A}.}
    \label{fig:pion_transm}
\end{figure}

\section{Results}\label{results}

The ratio spectrum in Figure \ref{fig:2090-2093_ratio_0.1ang_5-20ang} reveals distinct features corresponding to the rest frame wavelengths of the \ion{Ne}{IX} resonance line (13.447 \AA), the Fe UTA absorption complex (16 – 17 \AA), the narrow \ion{O}{VII} radiative recombination continuum (RRC at the \ion{O}{VII} edge, 16.771 \AA), the \ion{O}{VII} absorption edge (16.771 \AA), and the \ion{O}{VII}/\ion{O}{VIII} absorption lines (18.627 \AA $\,$ and 18.973 \AA, respectively).
As the Ne emission lines are relatively narrow, they are likely stable over time, appearing to deviate from unity due to the ratio spectrum's normalization against a variable continuum. 
And the optical depths of \ion{O}{VII} Ly $\beta$, \ion{O}{VIII} Ly $\alpha$, \ion{Ne}{IX} resonance line are very large, which indicate that all of these absorption lines are saturated. 
In contrast, the Fe UTA absorption complex and the \ion{O}{VII} edge ($16$ – $17.5$ \AA\ in the observed frame) exhibit significant variability, serving as the primary diagnostic indicators of warm absorber variation. By fitting the ratio spectrum with a combination of positive and negative Gaussian components, we determine a 10 $\sigma$ significance for Fe UTA variation, consistent with findings by \cite{Krongold2005ApJ}.

Figure \ref{fig:pion_transm} demonstrates that component 5, with $\rm log \, \xi = 1.65$ and column density = $0.5 \times 10^{26} \rm \, m^{-2}$, is the main contributor to the Fe UTA complex and \ion{O}{VII} edge. 
The spectral variations observed in the ratio spectrum (Figure \ref{fig:2090-2093_ratio_0.1ang_5-20ang}) reflect an increase in optical depth at the long-wavelength end of the Fe UTA and the \ion{O}{VII} edge during the low-flux state, suggesting a response of the ionization state of the warm absorber.

\begin{figure}[!ht]
  \centering
  \includegraphics[width=\linewidth]{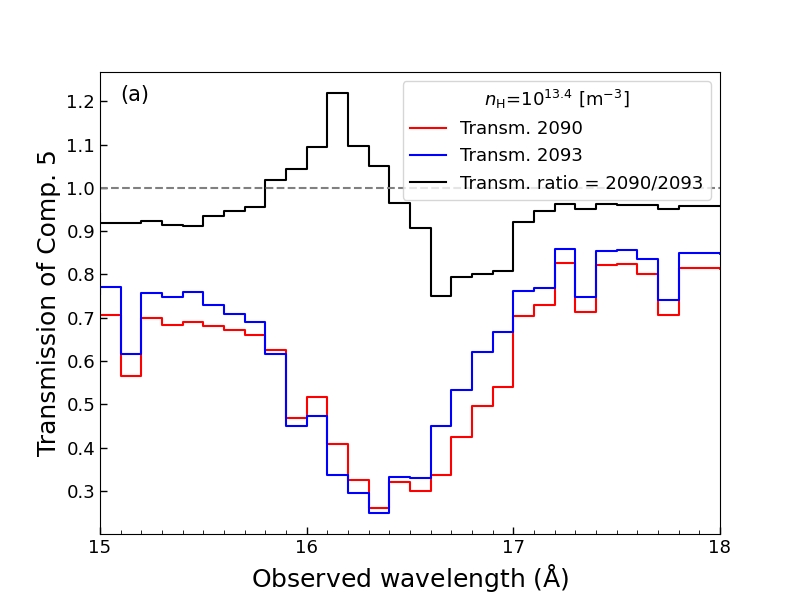}
  \includegraphics[width=\linewidth]{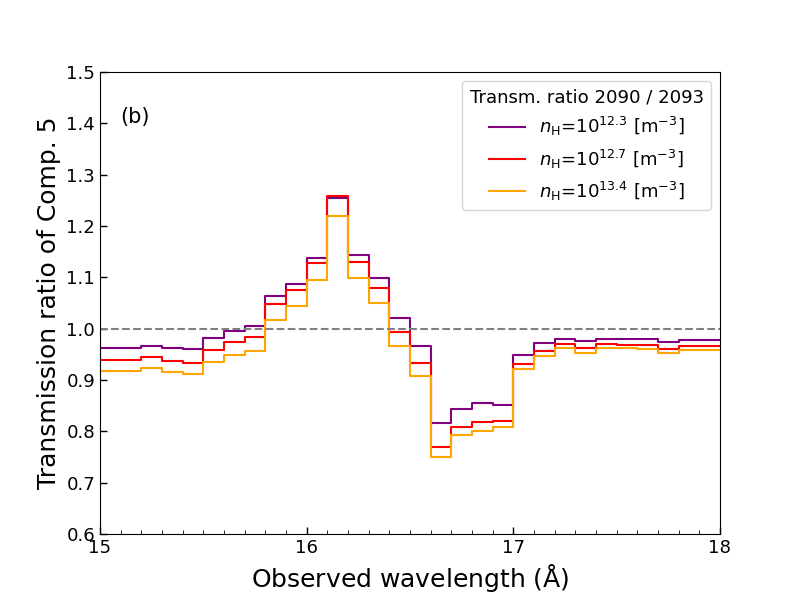}
  \includegraphics[width=\linewidth]{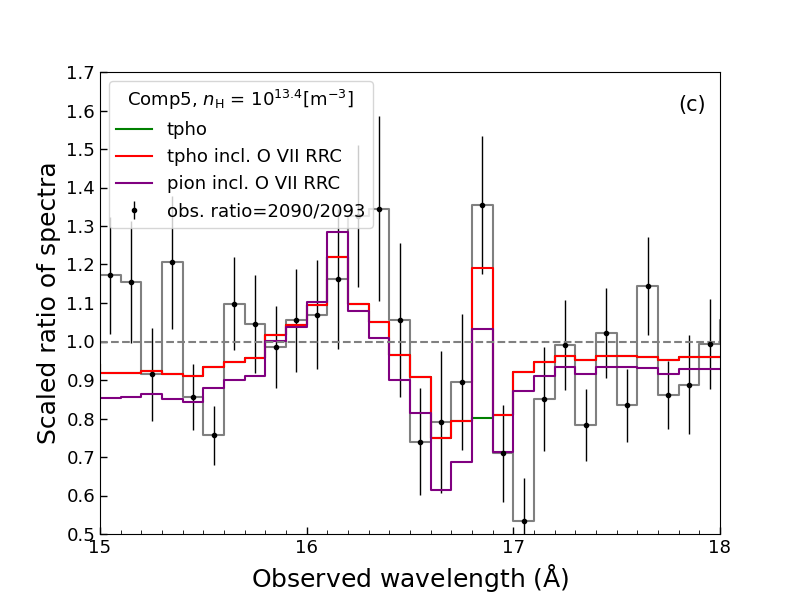}
  \caption{\texttt{tpho} models for component 5 in the $15 - 18 \, \rm \AA$ binned with $0.1 \rm \AA$.
  (a) Transmission of component 5 in high flux state (blue) and low flux state (red), transmission ratio (black) from low flux state over high; 
  (b) Transmission ratio of component 5 as a function of  increasing density.
  (c) The observed ratio (black) was fitted well by the \texttt{tpho} model with a density of $10^{13.4} \rm m^{-3}$ including the \ion{O}{VII} RRC feature (red). 
  The local scaling factor of 2.21 was calculated from the counts in the $9$ - $14$ \AA\ range. 
}
  \label{fig:transm_ratio_obs_fit}
\end{figure}

Figure \ref{fig:transm_ratio_obs_fit} displays the results from the \texttt{tpho} calculations for component 5 and the best fit to the observed ratio of ObsID 2090 over 2093 within the local wavelength range of $15$ – $18$ \AA.
Panel (a) shows a clear transmission change from low-flux (red) to high-flux (blue) states at the same density, with the transmission ratio (black) revealing prominent peaks and dips exceeding $20\%$ within the $16$ – $17$ \AA $\,$ wavelength range. 

Panel (b) of Figure \ref{fig:transm_ratio_obs_fit} shows that calculations at lower densities (purple) produce weaker or negligible variations within the $16$ – $17.5$ \AA\ band, which is attributed to longer recombination timescales and reduced variability in ionization structure. In contrast, higher-density calculations (red and yellow lines) amplify the contrast between low- and high-flux states, with the greatest amplification occurring at a density of $10^{13.4} \rm \, m^{-3}$, the current limit of our computational capabilities.

Panel (c) of Figure \ref{fig:transm_ratio_obs_fit} demonstrates that a density of $10^{13.4} \rm \, m^{-3}$ is optimal for reproducing the observed UTA and \ion{O}{VII} edge variations, as suggested by the \texttt{tpho} calculations for component 5.
Figure \ref{fig:2090-2093_ratio_0.1ang_5-20ang} displays a peak in the ratio spectrum corresponding to the narrow \ion{O}{VII} RRC at 16.85 \AA\ in the observed frame. To interpret this feature, we assume that the RRC component remains constant over time, with the peak arising solely from continuum variation. 
This assumption is reasonable, as the RRC likely originates from the narrow line region, which is characterized by low density (estimated around $~ 10^{10} \, \rm m^{-3}$, \citealp{Davies2020MNRAS}) and thus has a recombination timescale of approximately one year (see Table 4 of \citealt{Li2023A&A}).
In panel (c), the red line representing the model ratio under this assumption agrees well with the observed data. 
The green line showing the \texttt{tpho} calculation without adding \ion{O}{VII} RRC ratio gives a poorer fit.
Additionally, we present the \texttt{pion} model calculation (purple), which closely matches the \texttt{tpho} results at a density of $10^{13.4} [\rm m^{-3}]$ within the Fe UTA band. However, slight discrepancies are noted around the \ion{O}{VII} edge band, with the \texttt{pion} model displaying a slightly lower continuum level than \texttt{tpho}. This suggests additional continuum absorption in the \texttt{pion} calculation, likely arising from oxygen within this wavelength range. We will explore this aspect in greater detail in the subsequent discussion.

In addition to component 5, component 6 might also partially contribute to the Fe UTA and \ion{O}{VII} edge structure (see Fig. \ref{fig:pion_transm}). 
However, our \texttt{tpho} calculations reveal that variations in component 6 are relatively minor, showing changes of less than $10\%$, especially when compared to the more substantial variability predicted for component 5.

\begin{figure}
    \centering
    \includegraphics[width=\linewidth]{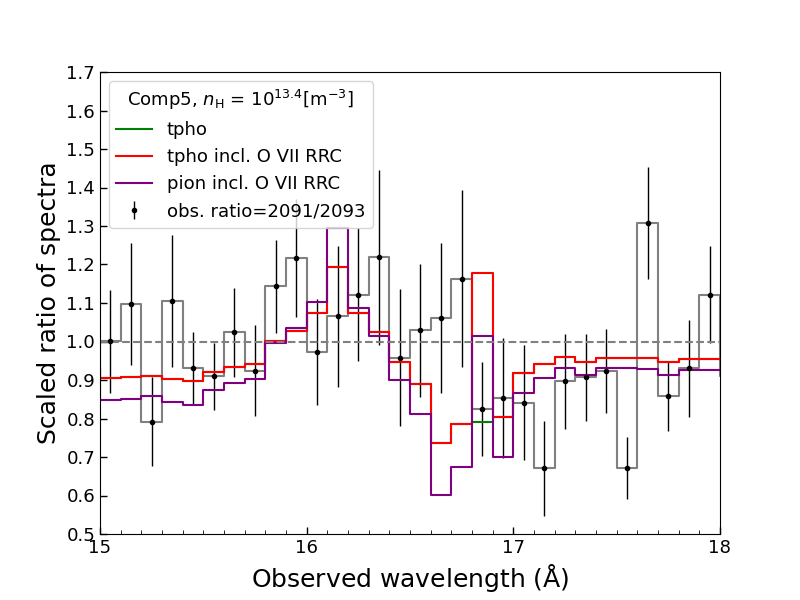}
    \includegraphics[width=\linewidth]{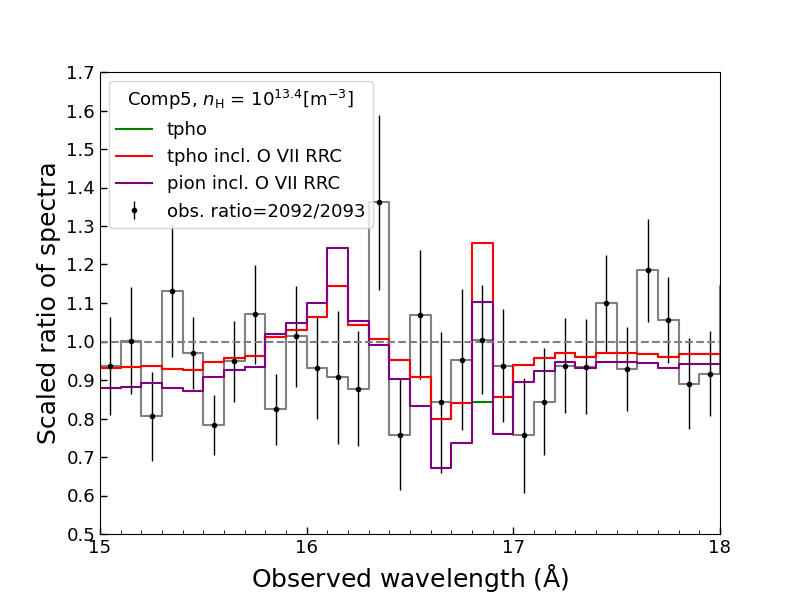}
    \includegraphics[width=\linewidth]{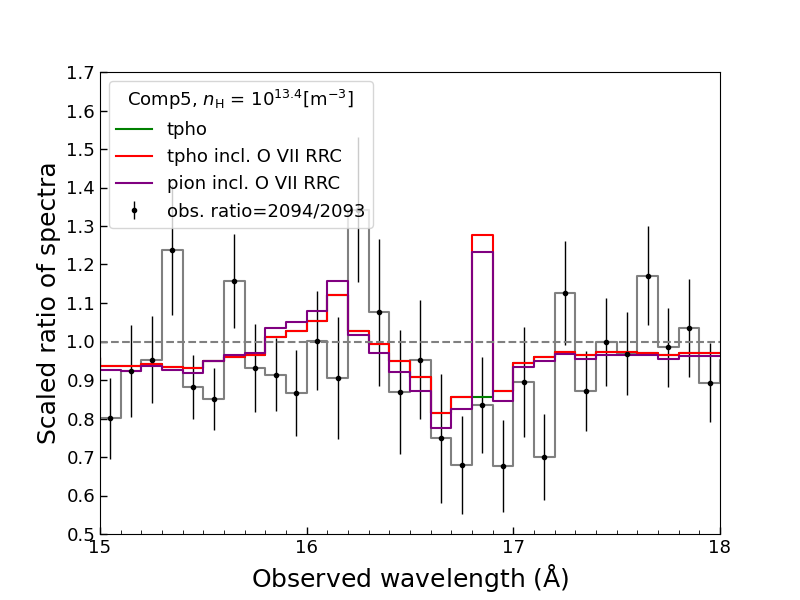}
    \caption{Best fitting of obs. ratio with model ratio for groups of ObsID 2091/2093, ObsID 2092/2093, ObsID 2094/2093, corresponding to the local scaling factor of $2.30$, $1.87$ and $1.66$ respectively. 
    The same approach was used as shown in Fig \ref{fig:transm_ratio_obs_fit}.
    }
    \label{fig: obs_ratio_model_other_3_groups}
\end{figure}

To further constrain the density of component 5, we conducted a joint analysis of the ratio spectra from all four low-flux observations (ObsIDs 2090, 2091, 2092, and 2094) relative to ObsID 2093 within the $15-18 \, \rm \AA$ range.
Figure \ref{fig: obs_ratio_model_other_3_groups} presents the ratio spectra (in black), fitted with the Component 5 model (in green), the Component 5 model including RRC (in red) with a density of $10^{13.4} \, \rm m^{-3}$, and the \texttt{pion} model (in purple).
Unlike ObsID 2090, the other three ObsIDs do not exhibit strong variation in ratio structure. Specifically, for ObsID  2094, the \texttt{tpho} model ratio for component 5 aligns well with the \texttt{pion} model, suggesting that the observed variations are within the expected range at this density level.

\begin{figure}[!htb]
   \begin{minipage}{\linewidth}
     \centering
     \includegraphics[width=1.1\linewidth]{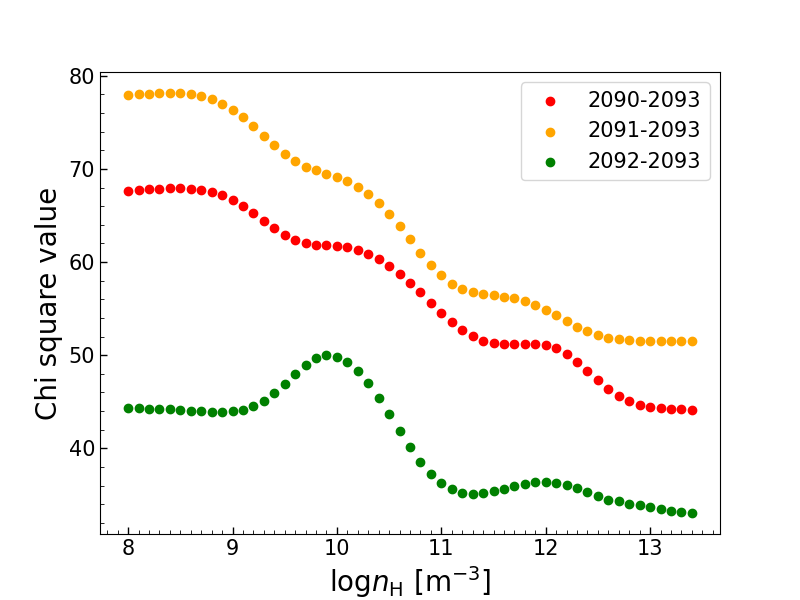}
     \caption{Chi-square values computed from a comparison of observed spectral ratios with model ratios computed with the \texttt{tpho} model. 
     }
     \label{Fig: 4 groups chi-square ratio}
   \end{minipage}\hfill
   \begin{minipage}{\linewidth}
     \centering
     \includegraphics[width=1.1\linewidth]{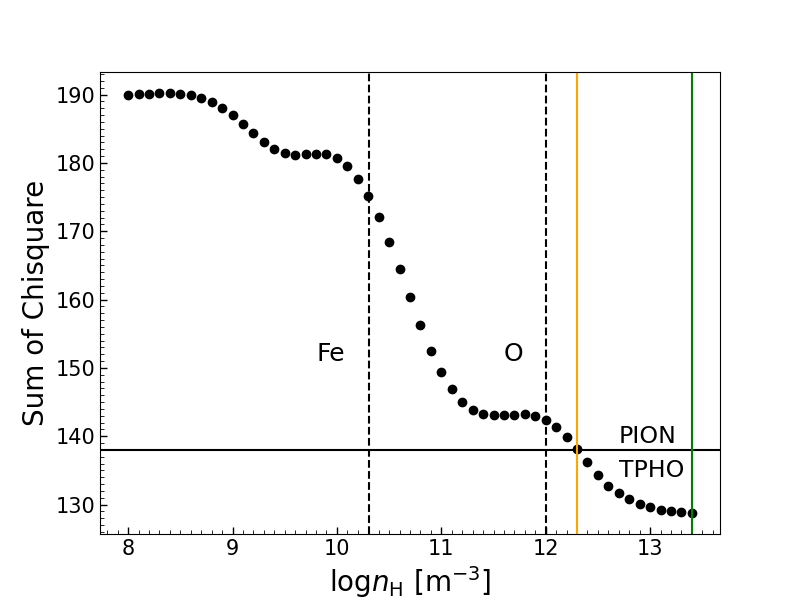}
     \caption{Sum of chi-square values from Fig. \ref{Fig: 4 groups chi-square ratio} vs. density (black dots).
     The minimum chi-square value of $128.76$ is obtained for a density of $10^{13.4} \, \rm m^{-3}$ (green line).
     The chi-square value for fits with the \texttt{pion} model is shown as a solid horizontal line, which is higher than the \texttt{tpho} value at the $3 \, \sigma$ confidence level (orange line).
     The vertical dashed lines indicate the densities of the outflow corresponding to a recombination time scale of $10^{5} \, \rm s$ for oxygen and iron.
     }
     \label{Fig: summed chi-square ratio}
   \end{minipage}
\end{figure}

For component 5 ($\rm log \, \xi = 1.65$), we performed a series of \texttt{tpho} calculations across a density range from $10^8$ to $10^{15} \, \rm m^{-3}$. 
For each density, we calculated a chi-square value by comparing the model predictions to the observed ratio spectra (Figure \ref{Fig: 4 groups chi-square ratio}).
Because the sampling of the input light curve for ObsID 2094 is sparse ($2\times10^{5} \, \rm s$) especially in last two weeks, it cannot give a reliable constraint for densities $\geq 10^{12} \, \rm m^{-3}$ for oxygen, densities $\geq 10^{10.3} \, \rm m^{-3}$ for iron corresponding to the recombination timescale of $2\times10^{5} \, \rm s$ for that density, separately (see vertical dashed lines of Figure \ref{Fig: summed chi-square ratio} and refer to \citealp{Li2023A&A}).
For the remaining observations, the sampling time is much shorter, of the order of $10^{4} \, \rm s$ (see Fig. \ref{fig:6 lc for tpho}), and we get reliable predictions for densities up to $10^{} \, \rm m^{-3}$.
Again, for higher densities the sampling uncertainties lead to too high uncertainties in the predicted transmission.
We combine the results for the three comparisons of Fig. \ref{Fig: 4 groups chi-square ratio} in Fig. \ref{Fig: summed chi-square ratio}.
The density corresponding to the minimum chi-square value of $128.76$, which is $10^{13.4} \, \rm m^{-3}$, represents the best-fit solution confidently.
At the $3 \sigma$ confidence level, we find that the density is higher than $10^{12.3} \, \rm m^{-3}$.
The conservative lower limit by the \texttt{tpho} model calculation indicates that the assumption of photoionization equilibrium (\texttt{pion} in the horizontal solid line with $\chi^{2} = 138$) is less accurate than the \texttt{tpho} predictions at the $3.1 \, \sigma$ confidence level ($\chi^{2} = 137$).
Additionally, Our lower limit is at least two orders of magnitude higher compared to the estimate based on photoionization equilibrium by \cite{Krongold2005ApJ}, who derived a density of $> 1 \times 10^{10} \, \rm m^{-3}$.

\section{Discussion}\label{discussion}

Through our analysis of the extensive Chandra/HETGS data set from the 2001 NGC 3783 campaign, we have identified significant spectral variability within the $16-17 \, \rm \AA$ range, with the most notable changes occurring on a monthly timescale. This finding indicates that the flux variation is likely driven by long-term fluctuations (on the order of month or more), which influence the plasma state of the warm absorber.
Time-dependent photoionization modeling using \texttt{tpho}, incorporating the RXTE light curve with a daily binning, suggests that the observed variability is predominantly attributable to component 5 of the warm absorber ($\rm log \xi = 1.65$). This component requires a density of at least $2.0 \times 10^{12} \, \rm m^{-3}$ to account for the variability, thus constraining its location to within $0.27 \, \rm pc$.

Our results confirm the flux-dependent features reported by \cite{Krongold2005ApJ}. However, the density limits derived in our study are significantly higher than those presented in their work. 
This difference can likely be attributed to the advancements in our model, which employs the latest time-dependent photoionization calculations and updated atomic data. Additionally, our findings are consistent with the density range reported by \cite{Gu2023A&A}, who also utilized the \texttt{tpho} model but employed a fundamentally different approach that focused solely on spectroscopic fitting of the time-averaged spectrum.

\begin{figure}
    \centering
    \includegraphics[width=1.1\linewidth]{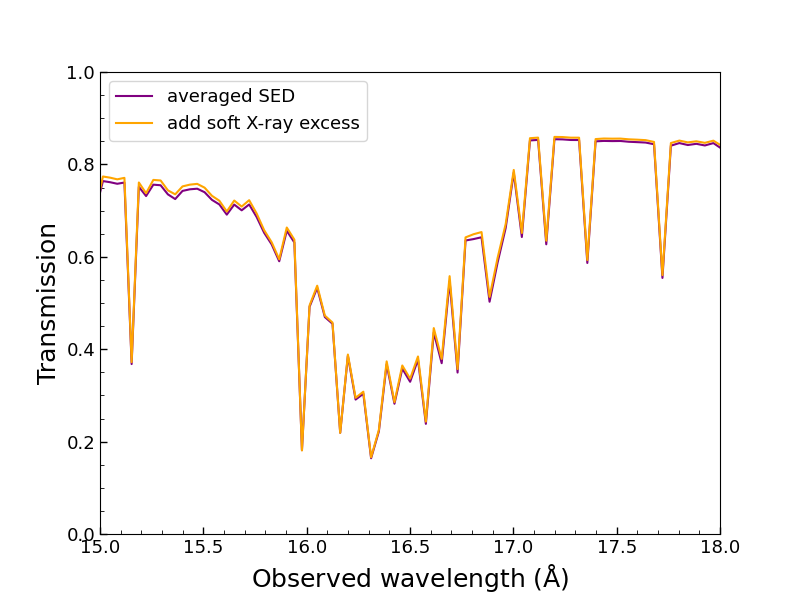}
    \caption{Transmission of component 5 with PION calculation by using different SED.
    Purple: averaged SED, orange curve: adding soft X-ray excess in wavelength of $10$ - $30$ \AA\ .}
    \label{fig:PION_with_different_SED}
\end{figure}
The soft excess reported by \cite{Netzer2003ApJ} and \cite{Krongold2005ApJ} could potentially introduce systematic uncertainties into our results, as the current \texttt{tpho} model does not account for changes in the SED. 
To assess the impact of the soft excess, we increased the flux in the $10-30 \, \text{Å}$ range by $20\%$ and reran the \texttt{tpho} calculation for component 5.  As shown in Figure \ref{fig:PION_with_different_SED}, the soft excess has a negligible effect on the time variability of the Fe UTA and O edge within the relevant wavelength range, particularly in our fitting range.

\cite{Netzer2003ApJ} concluded that the observed changes in the underlying continuum were attributed to the appearance (high flux state) and disappearance (low flux state) of the soft excess component. This conclusion is consistent with the observations from ObsIDs 2093 and 2090, as shown in Fig. \ref{fig:chandra 6 obs}.
The significant Fe UTA structure observed in ObsID 2093, following ObsID 2090, suggests that the Fe charge state has changed on a monthly timescale according to \texttt{tpho} calculations, allowing us to constrain the density of the low-ionization component of the warm absorbers.
Fig. \ref{fig:PION_with_different_SED} demonstrates that the soft X-ray excess does not significantly affect the Fe UTA feature, especially under the assumption of high density for the warm absorber component, as predicted by the \texttt{pion} model.

\begin{figure}
    \centering
    \includegraphics[width=1.1\linewidth]{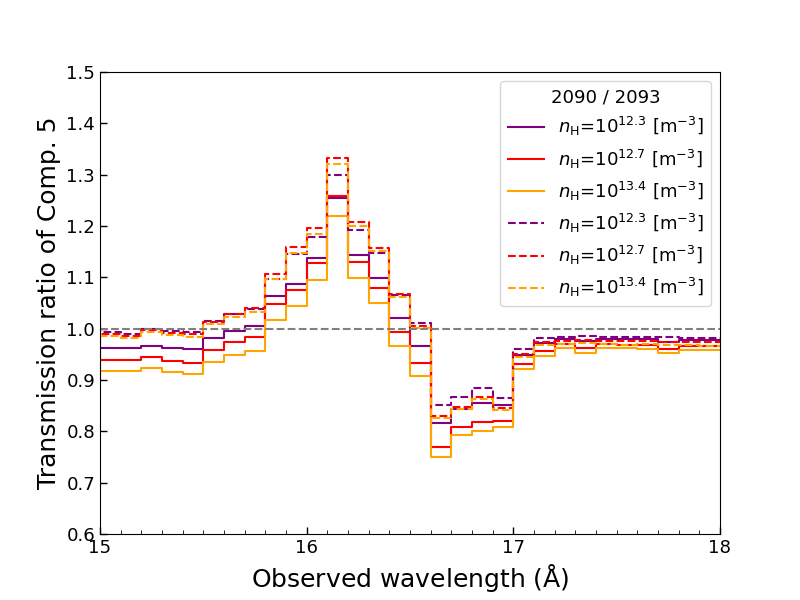}
    \caption{
    Transmission ratio of component 5 with the \texttt{tpho} model for different density in solid and dashed lines (including or excluding oxygen, respectively).
    }
    \label{fig:without oxygen}
\end{figure}

\begin{figure}
    \centering
    \includegraphics[width=1.1\linewidth]{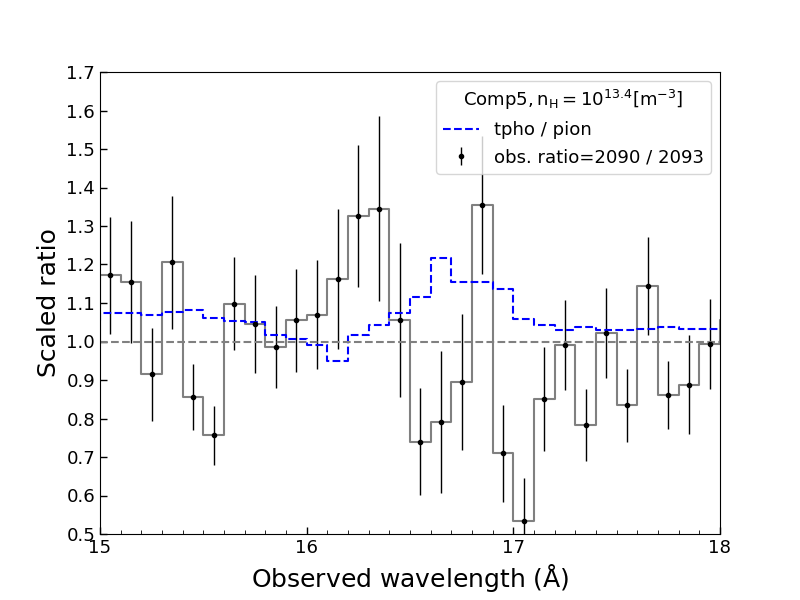}
    \caption{
    Transmission ratio of \texttt{tpho} over \texttt{pion} is represented by the blue line, observation ratio of ObsID 2090 to 2093 was illustrated in black curve. 
    }
    \label{fig:the ratio of tpho ratio to pion}
\end{figure}

As illustrated in Figure \ref{fig:transm_ratio_obs_fit}, component 5 exhibits absorption features primarily from the Fe UTA and \ion{O}{VII} edge at the relevant wavelengths.
Due to their distinct recombination timescales—approximately $10^3$ seconds for Fe and $10^5$ seconds for \ion{O}{VII} at a density of $10^{12} \, \text{m}^{-3}$—Fe and \ion{O}{VII} respond differently to source variations, with Fe reacting to more rapid fluctuations while \ion{O}{VII} primarily reflects longer-term changes.
To fully separate the spectral contributions from Fe and \ion{O}{VII}, a detailed analysis of spectral variations on hour- and day-level timescales would be ideal, though such data are not immediately available in this study.
However, an estimate of their relative contributions to the observed spectral ratio variation can still be obtained. By setting the oxygen column density to zero in the \texttt{tpho} calculation, we isolate the Fe UTA feature, represented as the dashed line, and compare it to the combined Fe and \ion{O}{VII} calculation, shown as the solid line in Figure \ref{fig:without oxygen}. 
The results suggest that Fe UTA absorption predominantly drives the variation in the $16 - 16.6 \, \rm \AA$ range, while the \ion{O}{VII} edge contribution is secondary.

The comparison between the \texttt{tpho} and \texttt{pion} models, as shown in panel (c) of Figure \ref{fig:transm_ratio_obs_fit} reveals subtle differences between the out-of-equilibrium and equilibrium assumptions, particularly in the $16.7 - 17 \, \rm Å$ range, where the \ion{O}{VII} edge contributes.
By calculating the ratio of the \texttt{tpho} to the \texttt{pion} transmission ratios, shown as the blue line in Figure \ref{fig:the ratio of tpho ratio to pion}, we observe that in the Fe UTA band, the \texttt{tpho} model with a density of $10^{13.4} \, \rm m^{-3}$ matches closely with \texttt{pion}. However, in the $16.7 - 17 \, \rm Å$ range, where the \ion{O}{VII} edge impacts absorption, \texttt{tpho} shows a 10\% difference from \texttt{pion}.

Considering this, and noting the slight decrease in continuum absorption in the \texttt{pion} model in Figure \ref{fig:transm_ratio_obs_fit}, we hypothesize that in ObsIDs 2090 and 2093, Fe has likely reached equilibrium while oxygen has not. This is due to the fact that the recombination timescale for Fe (100 seconds at $10^{13.6} \, \rm m^{-3}$) is considerably shorter than the light curve sampling interval, allowing Fe to reach equilibrium quickly (also refer to Figure \ref{fig:6 lc for tpho}). 
In contrast, oxygen, with a recombination timescale of approximately $10^4 \, \rm s$ at the same density, aligns more closely with the sampling binsize, causing its response to lag behind the flux variations observed.
Consequently, given the gaps between observations, oxygen could serve as an insightful diagnostic tool to probe higher-density plasmas, where recombination timescales more effectively reveal out-of-equilibrium effects (see Fig. \ref{Fig: summed chi-square ratio}). This further emphasizes the importance of considering both density and recombination timescales in characterizing plasma conditions in variable environments.


The upper limit of the density $n_{\rm H, upp}$ for an escaping wind is derived from the assumption that the outflowing velocity $v_{\rm out}$ is larger than or equal to the escape velocity $v_{\rm esc} = \sqrt{2GM_{\rm BH} / r}$.
This yields 
\begin{equation}
    n_{\rm H, upp} = \frac{L_{\rm ion} {v_{\rm out}^4}}{{4 (G M_{\rm BH})^2} \xi} \, 
\end{equation}
by substitute escaping velocity into Eq. \ref{eq1}.
The lower limit density $n_{\rm H}$ of component 5 from our present \texttt{tpho} modelling is larger than $ n_{\rm H, upp}$ (see Figure 8 of \citealp{Li2023A&A}), which indicates that the velocity of component 5 remains below the escape velocity.
Therefore, this velocity discrepancy suggests that component 5 may be a failed wind, unable to reach velocities sufficient to escape the gravitational pull of the SMBH, thus remaining bound to the system.

Based on the density constraints shown in Figure \ref{Fig: summed chi-square ratio}, we can constrain the distance $r$ of the low-ionization warm absorber component using Eq. \ref{eq1}.
Inserting log$\rm \xi$ = 1.65, $L_{\rm ion} = 6.36 \times 10^{36} \, \rm W$ (\citealp{Mehdipour2017A&A}), at the $3 \sigma$ confidence level,  
the distance of component 5 is limited to less than $0.27 \, \rm pc$. 
The derived distance is notably closer to the central SMBH than previous estimates, such as the $\leq$ 6 pc derived by \cite{Krongold2005ApJ}.
This closer proximity aligns well with density values inferred from other variability studies, such as the upper bounds suggested by \cite{Behar2003ApJ} based on XMM-Newton grating spectra analysis. 
Furthermore, this finding of location is consistent with the modelling by \cite{Chelouche2005ApJ}, who predicted that hot X-ray absorbers likely originate from within $1 \, \rm pc$ by analyzing thermodynamic conditions in the outflowing structures.


Combining the results from the \cite{GRAVITY_Collaboration_2021A&A}, the upper distance limit for component 5 (0.27 pc) is smaller than the coronal line region (CLR, 0.4 pc) but larger than the hot dust (0.14 pc) and broad line region (BLR, 16 light-days) as measured by Br $\gamma$. This is consistent with the predicted scenario of a low-ionization radiatively-driven wind, which fails to escape and returns to the disk \citep{Gallo2023arXiv}.

This "failed wind" component could provide a physical framework for interpreting various phenomenological components of AGN \citep{Giustini2021IAUS}. For instance, the inner failed wind component might offer a physical interpretation for key AGN features, such as the high-ionization BLR, the obscuring material, and the X-ray corona. Additionally, the failed wind solutions describing the inner accretion and ejection flows of AGN could help assess whether they significantly alter the physical and geometrical structure of the innermost accretion flow around highly accreting SMBHs.

In any model that predicts the launching mechanism and the impact of outflows on the surrounding medium, it is essential to determine the precise distance at which the outflow is launched. Time-dependent photoionization modelling, which incorporates the variability of the warm absorber, offers an advantage in accurately determining the gas density. High-sensitivity, high-resolution calorimeters (such as Resolve onboard XRISM, \citealp{Tashiro2020SPIE11444}, and X-IFU prepared for the future Athena X-ray observatory, \citealp{Barret2018SPIE}) will provide significant advancements, especially in the study of variable warm absorbers and in determining their gas densities.

\section{Conclusions}\label{sect:5}

In this study, we revisited the unobscured-state spectrum of NGC 3783 using archival Chandra/HETGS data from the 2001 campaign, investigating spectral variability over timescales of several months by comparing ratio spectra across varying flux levels. Significant changes were observed in the ratio spectrum over week-to-month timescales, corresponding to a flux variation factor of $1.4 - 2$. This variability, detected at a significance level greater than 10 $\sigma$ at maximum flux, points to notable response in the ionizing continuum.
To further explore these changes, we applied time-dependent photoionization modelling with \texttt{tpho} (SPEX) and found that the low-ionization component ($\rm log \xi = 1.65$) exhibits a pronounced sensitivity to flux variations in the Fe UTA band. 
We constrained the density of this low-ionization component to be higher than $10^{12.3}$ $ \rm m^{-3}$ at the $3 \sigma$ confidence level. This density places the location of the component within $0.27$ pc of the central source.
Our results suggest that this low-ionization component may represent a failed wind candidate, as it lacks sufficient momentum to escape the gravitational influence of the SMBH, instead remaining bound within the system. This finding contributes to a refined understanding of the dynamics and structure of AGN outflows in NGC 3783.


%



\begin{acknowledgements}
      We thank the anonymous referee for his/her constructive comments.
      C.L. acknowledges support from Chinese Scholarship Council (CSC) and Leiden University/Leiden Observatory, as well as SRON.
      SRON is supported financially by NWO, the Netherlands Organization for Scientific Research.
      C.L. thanks Elisa Costantini for the discussions of distance scale of outflowing wind.
\end{acknowledgements}

%
%

\bibliographystyle{aa} 
\bibliography{papercite.bib}

\end{document}